\documentclass[utf8]{frontiersSCNS} 

\usepackage{url,hyperref,lineno,microtype,subcaption}
\usepackage[onehalfspacing]{setspace}
\usepackage{gensymb}
\usepackage{mathrsfs}

\def\keyFont{\fontsize{8}{11}\helveticabold }
\def\firstAuthorLast{Liu {et~al.}} 
\def\Authors{Bingchuan Liu\,$^{1}$, Xiaoshan Huang\,$^{1}$,Yijun Wang\,$^{2}$, Xiaogang Chen\,$^{3}$, and Xiaorong Gao\,$^{1,*}$}
\def\Address{$^{1}$Department of Biomedical Engineering, Tsinghua University, Beijing, 100084, China\\
$^{2}$State Key Laboratory on Integrated Optoelectronics,
Institute of Semiconductors, Chinese Academy of Sciences, Beijing, 100083, China,\\
$^{3}$Institute of Biomedical Engineering, Chinese Academy of Medical Sciences and Peking Union Medical College, Tianjin, 300192, China}

\def\corrAuthor{Corresponding Author}

\def\corrEmail{gxr-dea@tsinghua.edu.cn}

\begin{document}
\onecolumn
\firstpage{1}

\title[BETA: A Large Benchmark Database Toward SSVEP-BCI Application]{BETA: A Large Benchmark Database Toward SSVEP-BCI Application} 
\author[\firstAuthorLast ]{\Authors} 
\address{} 
\correspondance{} 

\extraAuth{}

\maketitle

\begin{abstract}

\section{}
Brain-computer interface (BCI) provides an alternative means to communicate and it has sparked growing interest in the past two decades. Specifically, for Steady-State Visual Evoked Potential (SSVEP) based BCI, marked improvement has been made in the frequency recognition method and data sharing. However, the number of pubic database is still limited in this field. Therefore, we present a \textbf{BE}nchmark database \textbf{T}owards BCI \textbf{A}pplication (BETA) in the study. The BETA database is composed of 64-channel Electroencephalogram (EEG) data from 70 subjects performing a 40-target cued-spelling task. The design and acquisition of BETA is in pursuit of meeting the demand from real-world applications and it can be used as a test-bed for these scenarios. We validate the database by a series of analysis and conduct the classification analysis of eleven frequency recognition methods on BETA. We recommend to use the metric of wide-band Signal-to-noise ratio (SNR) and BCI quotient to characterize the SSVEP at the single-trial and population level, respectively. The BETA database can be downloaded from the website http://bci.med.tsinghua.edu.cn/download.html.

\tiny
 \keyFont{ \section{Keywords:}  Brain-computer interface (BCI), Steady-state visual evoked potential (SSVEP), Electroencephalogram (EEG), Public database, Frequency recognition, Classification algorithms, Signal-to-noise ratio (SNR)} 
\end{abstract}

\section{Introduction}
Brain-Computer Interface (BCI) provides a new way of brain interaction with the outside world by measuring and converting brain signals to external commands without the involvement of peripheral nervous system (\cite{wolpaw2002brain}). The technology of BCI has considerable scientific significance and application prospect, especially on rehabilitation (\cite{ang2013brain-computer,lebedev2017brain-machine}) and as an alternative access method for the physically disabled (\cite{pandarinath2017high,gao2003a}). Steady-State Visual Evoked Potential (SSVEP) is a stable neural response elicited by periodic visual stimuli and the frequency tagging attribute can be leveraged in BCI (\cite{norcia2015the,cheng2002design}). Among a variety of BCI paradigms, SSVEP based BCI (SSVEP-BCI) has gained widespread attention due to its non-invasiveness, high signal-to-noise ratio (SNR) and high information transfer rate (ITR) (\cite{bin2009an,chen2015high-speed}). Generally, the high-speed performance of BCI is accomplished by a multi-target visual speller, which achieves a reportedly average online ITR of 5.42 bit per second (bps) (\cite{nakanishi2018enhancing}). Besides, the ease of use and significantly lower rate of BCI illiteracy (\cite{lee2019eeg}) make it a promising candidate for real-world applications.
\par To push the boundary of BCI toward the goal, rapid progress has been made to facilitate frequency recognition of SSVEP (see review (\cite{zerafa2018train})). Based on whether a calibration or training phase is required for the extraction of spatial filters, the signal detection methods can be categorized into supervised methods and training-free methods. The supervised methods exploit an optimal spatial filter by a training procedure and achieve the state-of-the-art classification performance in SSVEP-based BCI (\cite{nakanishi2018enhancing, wong2019learning}). These spatial filters or projection direction can be learned by exploiting individual training template (\cite{bin2011a}), reference signal optimization (\cite{zhang2013l1-regularized}), inter-frequency variation (\cite{yin2015a}), ensemble reference signals (\cite{nakanishi2014a,chen2015high-speed}) in the framework of canonical correlation analysis (CCA). Recently, task-related components (\cite{nakanishi2018enhancing}) and multiple neighboring stimuli (\cite{wong2019learning}) are utilized to derive spatial filters to further boost the discriminative power of the learned model. On the other hand, training-free methods perform feature extraction and classification in one shot without the training session in online BCI. This line of work usually use a sinusoidal reference signal and the detection statistics can be derived from canonical correlation (\cite{bin2009an}) and its filter-bank form (\cite{chen2015filter}), noise energy minimization (\cite{friman2007multiple}), synchronization index maximization (\cite{zhang2014multivariate}), and additional spectral noise estimation (\cite{Abu2016Advancing}), etc.
\par Along with the flourish of frequency recognition methods, continuous efforts have been devoted to share SSVEP database (\cite{bakardjian2010optimization,kolodziej2015new,kalunga2016online,kwak2017convolutional,icscan2018steady}) and contribute to public SSVEP database (\cite{wang2017a,lee2019eeg,choi2019multi}). \cite{wang2017a} benchmarked a 40-target database comprising 64-channel 5-s SSVEP trial for 35 subjects, each of which performed an offline cue-spelling task in 6 blocks. Recently, \cite{lee2019eeg} released a larger database of 54 subjects performing a 4-target offline and online task and 62-channel 4-s SSVEP data were obtained with 50 trials per class. \cite{choi2019multi} also provided a 4-target database and 6-s SSVEP data with physiological data are collected from 30 subjects at 3 different frequency bands (low: 1--12 Hz; middle: 12--30 Hz; high: 30--60 Hz) across two days. Up to date, the number of public databases in the SSVEP-BCI community is still limited compared to other domains, e.g. computer vision, where a growing number of databases play a critical role on the development of the disciplinary (\cite{russakovsky2015imagenet}). Compared to other BCI paradigms, e.g. motor imagery BCI, the databases for SSVEP-BCI are also scarce (\cite{choi2019multi}). Therefore, more databases are necessitated in the realm of SSVEP-BCI for the design and evaluation of methods.
\par To this end, we present a large \textbf{BE}nchmark database \textbf{T}owards SSVEP-BCI \textbf{A}pplication (BETA) in this study. The BETA database is composed of 70 subjects in a cued-spelling task. As an extension of the benchmark database (\cite{wang2017a}), the number of targets is 40 and the frequency range is also from 8 to 15.8 Hz. A key feature of the BETA database is that it's tailored for real-world applications. Different from the benchmark database, the BETA is collected outside the laboratory setting of the electromagnetic shielding room. Since it is imperative to reduce the calibration time from a practical perspective, the number of blocks decreases to 4 instead of 6 in the benchmark. A QWERT virtual keyboard is presented in flicker to better approximate the conventional input device and enhance user experience. To the best of our knowledge, the BETA database has the largest number of subjects for SSVEP-BCI to date. Since a larger database captures the inter-subject variability, the BETA database makes it possible to reflect a more realistic EEG distribution and potentially meet the demands of real-world BCI applications.
\par The remaining of the paper is organized as follows. First, we detail the procedure of data acquisition and curation in Section 2. Then we describe the data record and availability in Section 3. Third, we perform data validation and then evaluate algorithms by comparing eleven frequency recognition methods on BETA. Finally, Section 4 concludes and we discuss additional findings from the database.
\section{Materials and Methods}
\subsection{Participants}
Seventy healthy volunteers (42 males and 28 females) with a mean age of $25.0\pm7.97$ (mean $\pm$ std, ranging from 9 to 64 years) participated in the study. All participants had normal or corrected to normal vision and gave written informed consent before the experiment (for participants under 16 years old the consent was signed by their parents). The study was carried out in accordance with the Declaration of Helsinki and the protocol was approved by the ethics committee of Tsinghua University (No. 20190002).
\subsection{Recruitment and Inclusion Criteria}
Participants were recruited on a national scale to take part in the Brain-Computer Interface 2018 Olympics in China. The competition was hold to contest and award individuals with high performance of BCI (SSVEP, P300 and Motor Imagery). The seventy participants are those who participated in the second round of the contest  (SSVEP-BCI track) and all are not naive to SSVEP-BCI. Before enrollment, participants were informed that the data would be used for non-commercial scientific research. Participants who conformed to the experimental rules in the first round and were available to the schedule of the contest were included for the second round. All the participants met the following criteria: (1) no history of epileptic seizures or other neuropsychiatric disorders (2) no attention-deficit or hyperactivity disorder (3) no history of brain injury or intracranial implantation.
\subsection{Visual Speller}
\begin{figure}[h!]
\begin{center}
\includegraphics[width=16cm]{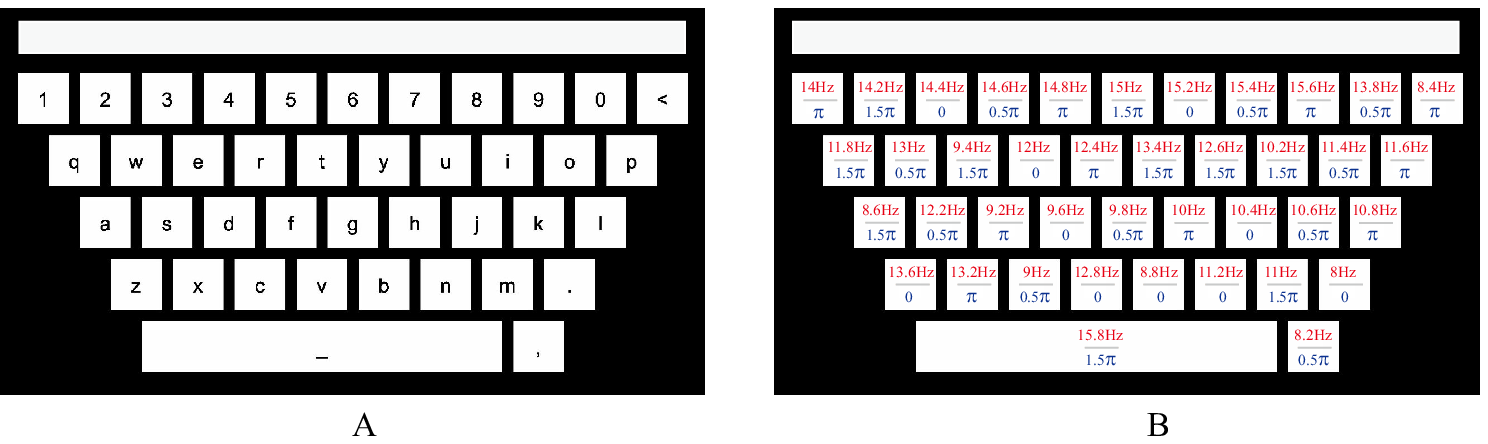}
\end{center}
\caption{The QWERT virtual keyboard for the 40-target BCI speller. \textbf{(A)} The layout resembles conventional keyboard with 10 numbers, 26 alphabets and 4 non-alphanumeric keys (dot, comma, backspace  \textless and space \_) aligned in 5 rows. The upper rectangle is designed for presenting the input character. \textbf{(B)} The frequency and initial phase for each target encoded in the joint frequency and phase modulation.}\label{fig:1}
\end{figure}

This study designed a 40-target BCI speller for visual stimulation. To facilitate the user experience, the graphical interface was designed to resemble the traditional QWERT keyboard. The keyboard was presented on a 27-inch LED monitor (ASUS MG279Q Gaming Monitor, 1920$\times$1080 pixels) at a fresh rate of 60 Hz. As illustrated in Figure\ref{fig:1}A, 40 targets including 10 numbers, 26 alphabets and 4 non-alphanumeric keys (dot, comma, backspace \textless \hspace{3pt} and space \_) were aligned in 5 rows, with a spacing of 30 pixels. The stimuli had the dimension of 136$\times$136 pixels (3.1\degree$\times$3.1\degree) for the square and 966$\times$136 pixels (21\degree$\times$3.1\degree) for the space rectangle. The topmost blank rectangle was for result feedback.

A sampled sinusoidal stimulation method (\cite{manyakov2013sampled,Chen2014A}) was adopted to present visual flicker on the screen. In general, the stimulus sequence of each flicker can be generated by the following equation 
\begin{equation}
    s(f,\phi,i)=\frac{1}{2}\{1+sin[2\pi f(i/\rm{RefreshRate})+\phi]\}
\end{equation}
where $i$ indicates the frame index in the stimulus sequence, and $f$ and $\phi$ indicate the frequency and phase values of the encoded flicker using a joint frequency and phase modulation (JFPM) (\cite{chen2015high-speed}). The grayscale value of stimulus sequence ranges from 0 to 1, where 0 indicates dark and 1 indicates the highest luminance of the screen. For the 40 targets, the tagged frequency and phase values can be obtained by
\begin{equation}
\begin{split}
    f_k=f_0 + (k-1)\cdot\Delta f\\
    \Phi_k=\Phi_0 + (k-1)\cdot\Delta \Phi\\
\end{split}
\end{equation}
where the frequency interval $\Delta f$ is 0.2 Hz and phase interval $\Delta \Phi$ is 0.5 $\pi$. $k$ denotes the index from dot, comma and backspace, followed by a to z and 0 to 9, and space. In the study, $f_0$ and $\Phi_0$ is set 8 Hz and 0, respectively. The parameters for each target are illustrated in Figure\ref{fig:1}B. The stimulus was presented by MATLAB (MathWorks, Inc.) using Psychophysics Toolbox Version 3 (\cite{brainard1997the}).
\subsection{Procedure}
This study consisted of 4 blocks of online BCI experiment using a cued-spelling task. Each block was composed of 40 trials corresponding to 1 trial for each stimulus target in a randomized order. Trials began with a 0.5-s cue (a red square covering the target) for gaze shift, followed by flickering on all the targets, and ended with 0.5 s for rest. Participants are asked to avoid eye blink during the flickering. During the 0.5-s rest, result feedback (one of the recognized characters) was presented in the topmost rectangle after online processing by a modified version of the FBCCA method (\cite{chen2015filter}). The flickering lasted at least 2 s for the first 15 participants (S1 - S15) and at least 3 s (S16 - S70) for the remaining 55 participants. To avoid visual fatigue, there was a short break between two consecutive blocks.
\subsection{Data Acquisition}
64-channel EEG data were recorded by SynAmps2 (Neuroscan Inc.) according to the international 10-10 system. The sampling rate was set 1000 Hz and the passband of hardware filter was 0.15-200Hz. A built-in notch filter was applied to remove 50-Hz power-line noise. Event triggers were sent from the stimulus computer to the EEG amplifier and synchronized to the EEG data by a parallel port as an event channel. All impedance of the electrodes was kept below 10 k$\Omega$. The vertex electrode Cz was used as a reference. During the online experiment, nine parietal and occipital channels (Pz, PO3, PO5, PO4, PO6, POz, O1, Oz and O2) were selected for online analysis to provide feedback result. To record EEG data in real-world scenarios, the data were recorded outside the electromagnetic shielding room.
\subsection{Data Preprocessing}
The previous study demonstrates that SSVEP harmonics in this paradigm have the frequency range up to around 90 Hz (\cite{chen2015high-speed},\cite{chen2015filter}). Based on the finding, a band-pass filtering (zero-phase forward and reverse filtering using eegfilt in EEGLAB (\cite{delorme2004eeglab:}) between 3 Hz and 100 Hz was conducted to remove environmental noise. Epochs were then extracted from each block, comprising 0.5 s before the stimulus onset, 2 s (for S1 - S15) or 3 s (for S16 - S70) of stimulation, and a subsequent 0.5 s. The subsequent 0.5 s may contain SSVEP data if the duration of the trial is greater than 2 s (for S1 - S15) or 3 s (for S16 - S70). Since frequency resolution does not affect the classification result of SSVEP (\cite{nakanishi2017does}), all the epochs were then down-sampled to 250 Hz. 
\subsection{Metrics}
The quality of SSVEP data can be evaluated quantitatively by signal-to-noise ratio (SNR) analysis and classification analysis. In the SNR analysis, most of the previous studies (\cite{chen2015filter,chen2015high-speed,xing2018a}) adopted the narrow-band SNR metric. The narrow-band SNR (in decibels, dB) is defined as the ratio of spectral amplitude at stimulus frequency to the mean value of the ten neighboring frequencies (\cite{chen2015filter})
\begin{equation}
SNR=20log_{10}\frac{y(f)}{\sum^5_{k=1}[y(f-\Delta f\cdot k)+y(f+\Delta f\cdot k)]}
\end{equation}
where $y(f)$ is the amplitude spectrum at frequency $f$ calculated by Fast Fourier Transform (FFT), and $\Delta f$ is the frequency resolution.

Based on the narrow-band SNR, we use the wide-band SNR as a primary metric to better characterize wide-band noise and the contribution of harmonics to the signals. The wide-band SNR (in decibels, dB) is defined as follows
\begin{equation}
SNR=10log_{10}\frac{\sum^{k=Nh}_{k=1}P(k\cdot f)}{\sum^{f=f_s/2}_{f=0}P(f)-\sum^{k=Nh}_{k=1}P(k\cdot f)}
\end{equation}
where $Nh$ is the number of harmonics, $P(f_n)$ is the power spectrum at frequency $f$ and $f_s/2$ is the Nyquist frequency. In the wide-band SNR, the sum of power spectrum of multiple harmonics ($Nh=5$) is regarded as signals and the energy of full spectral band subtracted from the signals is taken as noise. 

Classification accuracy and information transfer rate (ITR) are widely used in the BCI community to assess the performance of different subjects as well as algorithms. The ITR (in bits per min, bpm) can be obtained by (\cite{wolpaw2002brain})
\begin{equation}
ITR=60\cdot(log_2M+Plog_2P+(1-P)log_2\frac{1-P}{M-1})/T    
\end{equation}
where $M$ is the number of classes, $P$ is the classification accuracy and $T$ (in seconds) is the average time for a target selection. The $T$ in the equation comprises gaze time and overall gaze shift time (e.g. 0.55 s in line with the previous studies (\cite{wang2017a,chen2015filter})). 
\section{Record Description}
The database is freely available at http://bci.med.tsinghua.edu.cn/download.html for scientific research and it is stored as MATLAB mat format. The database contains 70 subjects and each subject corresponds to a mat file. The name of subjects are mapped to indices from S1 to S70 for de-identification. Each file consists of a MATLAB structure array, with 4-block EEG data and its counterpart supplementary information as its fields.
\subsection{EEG Data}
EEG data after preprocessing are store as a 4-way tensor, with a dimension of channel $\times$ time point $\times$ block $\times$ condition. Each trial comprises 0.5-s data before the event onset and 0.5-s data after the time window of 2 s or 3 s. For S1-S15, the time window is 2 s and the trial length is 3 s, whereas for S16-S70 the time window is 3 s and the trial length is 4 s. Additional details about the channel and condition information can be found in the following supplementary information.
\subsection{Supplementary Information}
Eight supplementary information is comprised of personal information, channel information, frequency and initial phase associated to each condition, SNR and sampling rate. The personal information contains age and gender of the subject. For the channel information, a location matrix (64 $\times$ 4) is provided, with the first column indicating channel index, the second column and third column indicating the degree and radius in polar coordinates, and the last column indicating channel name. The SNR information contains the mean narrow-band SNR and wide-band SNR matrix for each subject, calculated in (3) and (4), respectively. The initial phase is in radius.
\section{Data Evaluation}
\subsection{Temporal, Spectral and Spatial Analysis}
\begin{figure}[h!]
\begin{center}
\includegraphics[width=16cm]{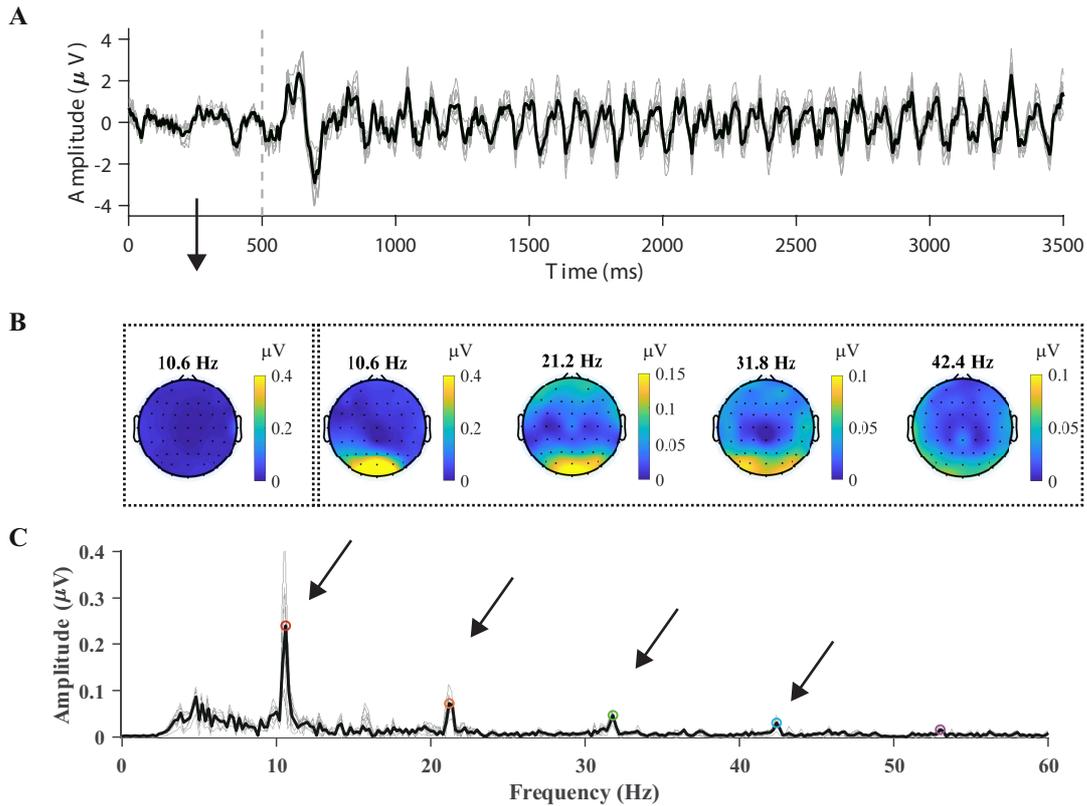}
\end{center}
\caption{Typical SSVEP features in temporal, spectral and spatial domain. \textbf{(A)} Time course of average 10.6-Hz SSVEP from nine parietal and occipital channels (Pz, PO3, PO5, PO4, PO6, POz, O1, Oz and O2). The dash line represents stimulus onset. \textbf{(B)} Topographic maps of SSVEP amplitudes at frequencies from fundamental signal (10.6 Hz) to fourth harmonic (21.2 Hz, 31.8 Hz and 42.4 Hz). The leftmost scalp map indicates the spectral amplitude at the fundamental frequency before stimulus. \textbf{(C)} The amplitude spectrum of SSVEP from the nine channels at 10.6 Hz. Up to 5 harmonics are visible from the amplitude spectrum. The averaged spectrum across channels is represented in the dark line in \textbf{(A)} and \textbf{(C)}.}\label{fig:2}
\end{figure}

To validate the data quality by visual inspection, nine parietal and occipital channels (Pz, PO3, PO5, PO4, PO6, POz, O1, Oz and O2) are selected and epochs are averaged with respect to the channels, blocks, and subjects. For consistency in data format, the subjects from S16 to S70 are chosen. Figure \ref{fig:2}A illustrates the averaged temporal amplitude for one stimulus frequency (10.6 Hz). After a delay (within 100-200 ms) at the stimulus onset, a steady-state and time-locked characteristic can be observed in the temporal sequence (Figure \ref{fig:2}A). Data between 500 ms and 3500 ms are extracted and padded with 2000-ms zeros to yield a 0.2-Hz spectral resolution in Figure \ref{fig:2}C. From the amplitude spectrum, the fundamental frequency (10.6 Hz: 0.266 $\mu V$) and three harmonics (21.2 Hz: 0.077 $\mu V$, 31.8 Hz: 0.054 $\mu V$, 42.4 Hz: 0.033 $\mu V$) are distinguishable from background EEG. Note that for high frequencies ($>$60Hz), both the harmonic signals and noise are small in amplitude due to the volume conduct effect (\cite{Broek1998Volume}) and are not shown here. 

\begin{figure}[h!]
\begin{center}
\includegraphics[width=12cm]{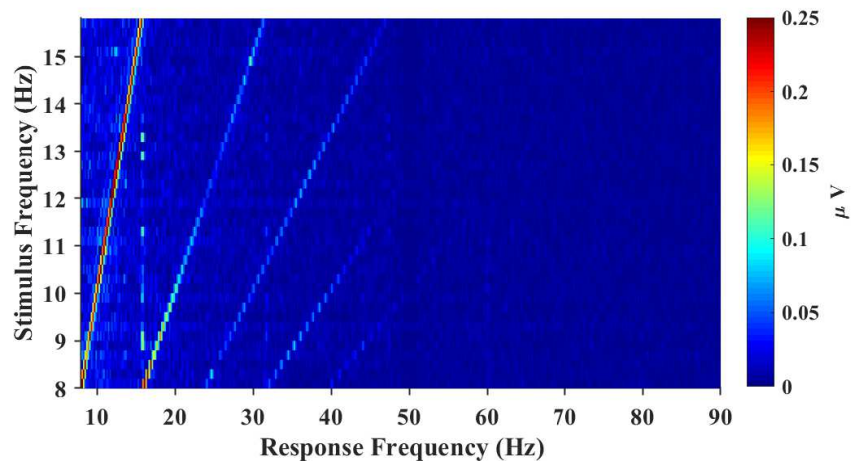}
\end{center}
\caption{The amplitude spectrum as a function of stimulus frequency (frequency range: 8--15.8 Hz; frequency interval: 0.2 Hz). The spectral response of SSVEP decreases rapidly as the number of harmonics increases and up to 5 harmonics are visible from the figure.}\label{fig:4}
\end{figure}

Figure \ref{fig:2}B illustrates topographic mappings of the spectrum at frequencies from fundamental signal to fourth harmonic. This indicates that fundamental and harmonic signals of SSVEP are distributed predominantly in the parietal and occipital regions. Frontal and temporal regions of the topographic maps also show an increase in the spectrum, which may represent noise or propagation of SSVEP oscillation from the occipital region (\cite{thorpe2007identification,liu2017effects}). To characterize response property of SSVEP, the amplitude spectrum as a function of stimulus frequency is plotted and illustrated in Figure \ref{fig:4}. From the amplitude spectrum, the spectral response of SSVEP decreases rapidly as the number of harmonics increases (up to 5 harmonics are visible). A dark line at 50-Hz response frequency results from notch filtering. A bright line at 15.8-Hz response frequency may be distractor stimulus from the SPACE target with a larger size.
\subsection{SNR Analysis}
\begin{figure}[h!]
\begin{center}
\includegraphics[width=16cm]{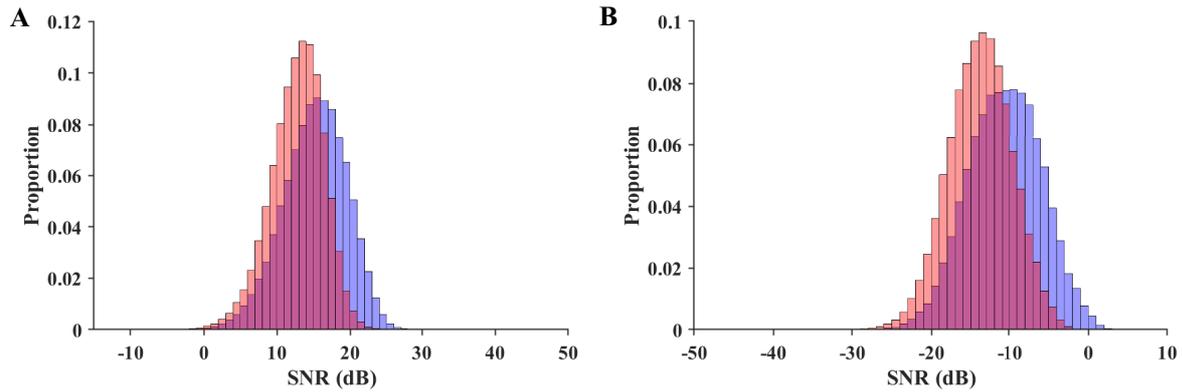}
\end{center}
\caption{Normalized histogram of narrow-band SNR \textbf{(A)} and wide-band SNR \textbf{(B)} for trials in the benchmark database and BETA. The red diagram indicates the BETA and the blue diagram indicates the benchmark database.}\label{fig:5}
\end{figure}

As a metric independent of different classification algorithms, SNR measures available stimulus-evoked components in SSVEP spectrum. For SNR analysis, we compare the BETA database with the benchmark database for SSVEP-based BCI (\cite{wang2017a}). Narrow-band SNR and wide-band SNR are calculated by (3) and (4) respectively for each trial. For a valid comparison, EEG in the benchmark database are band-pass filtered between 3 Hz and 100 Hz (eegfilt in EEGLAB) before epoching. Trials in this database are padded with zeros (3s for S1-S15, 2s for S16-S70) to yield a spectral resolution of 0.2 Hz. Figure \ref{fig:5} illustrates the normalized histogram of narrow-band SNR (A) and wide-band SNR (B) for trials in the two databases. For narrow-band SNR, the BETA database has a significantly lower SNR (4.019$\pm$0.018 dB) than that of benchmark database (8.185$\pm$0.024 dB), with a p-value $\textless 0.001$ (two-sample unpaired t-test). Similarly, the wide-band SNR in the BETA database (-13.758$\pm$0.013 dB) is significantly lower than that of benchmark database (-10.6287$\pm$0.017 dB), with a p-value $\textless 0.001$ (two-sample unpaired t-test). This is due in part to individual differences in SNR for the two studies and EEG is recorded outside the electromagnetic shielding room in the BETA database. The comparable results of the two SNRs also demonstrate the validity of the wide-band SNR metric that takes additional information of wide-band noise and harmonics into account. 

\begin{figure}[h!]
\begin{center}
\includegraphics[width=16cm]{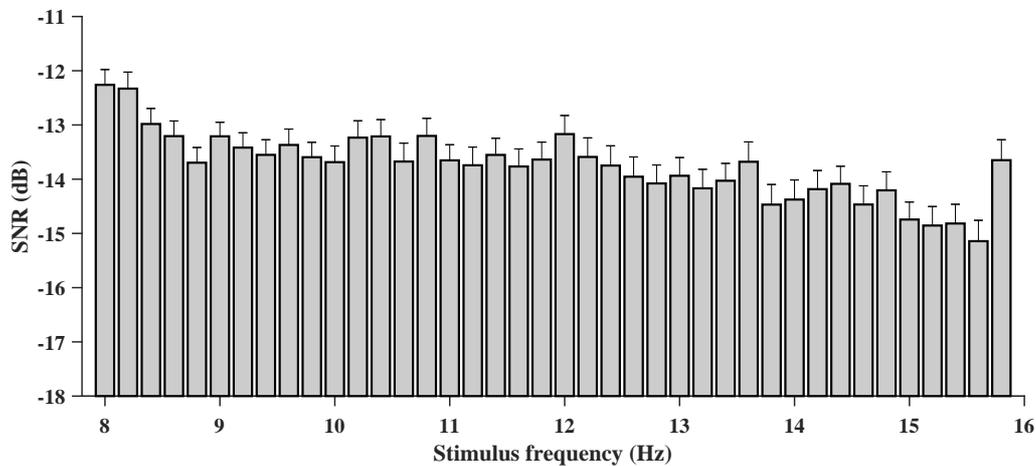}
\end{center}
\caption{The wide-band SNR corresponding to the 40 stimulus frequencies (from 8 Hz to 15.8 Hz with an interval of 0.2 Hz). A general tendency of decline in SNR can be observed as the stimulus frequency increases. The SNR is higher at 15.8 Hz presumably because the target has a larger shape of the region.}\label{fig:6}
\end{figure}

In addition, we analyze the characteristics of SNR with respect to each stimulus frequency. For the BETA database, the wide-band SNR is calculated by each zero-padded trial and the SNR associated with each condition is obtained by average per block and per person. Figure \ref{fig:6} illustrates the wide-band SNR corresponding to the 40 stimulus frequencies. In general, a tendency of decline in SNR can be observed as the stimulus frequency increases. For some stimulus frequencies, e.g. 11.6 Hz, 10.8 Hz, 12 Hz and 9.6 Hz, the SNR bumps up compared to their adjacent frequencies. Specifically, the SNR at 15.8 Hz is elevated by an average of 1.49 dB compared to 15.6 Hz, presumably due in part to the larger shape of the region for visual stimulation.
\subsection{Phase and Visual Latency Estimation}
\begin{figure}[h!]
\begin{center}
\includegraphics[width=16cm]{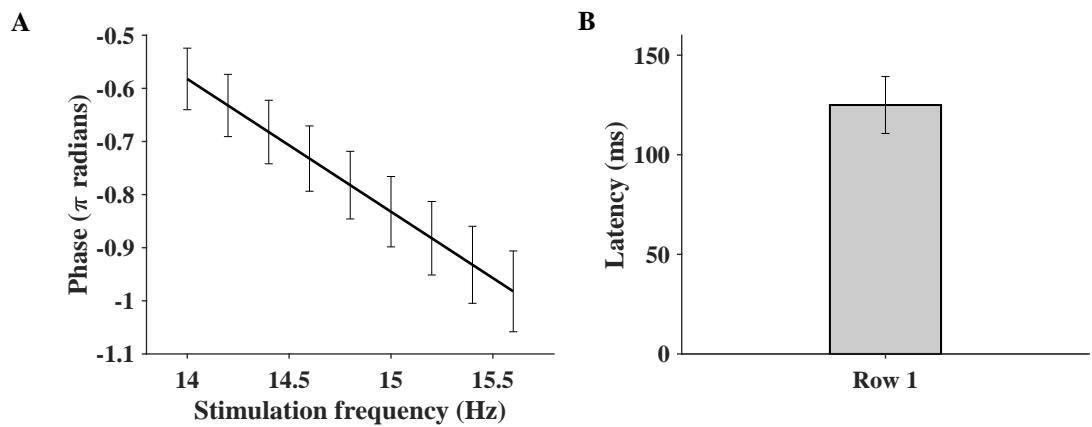}
\end{center}
\caption{The phase as a function of stimulus frequency \textbf{(A)} and the bar plot of estimated latencies \textbf{(B)}. SSVEP from Oz channel at nine consecutive stimulus frequencies (row 1 of the keyboard) is extracted for analysis. The errorbar indicates the standard deviation.}\label{fig:7}
\end{figure}

To further make a comparison with the benchmark database in the previous study (\cite{wang2017a}), we conduct an estimation of phase and visual latency in the BETA database. Nine consecutive stimulus frequencies in row 1 of the keyboard are selected and the SSVEP from Oz channel (70 subjects) is extracted for analysis. The procedure is performed according to the previous study (\cite{wang2017a}) using a linear regression between the estimated phase and stimulus frequency (\cite{di1999electrophysiological}). The visual latency for each subject can be estimated by the slope $k$ of the linear regression in the form
\begin{equation}
\rm{Latency}=-500\cdot \textit{k}
\end{equation}
Figure \ref{fig:7} illustrates the phase as a function of stimulus frequency and the bar plot of estimated latencies from (6). The mean estimated visual latency in this study is 124.96 $\pm$ 14.81 ms, which is close to 136.91 $\pm$ 18.4 ms in the benchmark database (\cite{wang2017a}) and approximates to 130 ms. Therefore, a 130-ms latency is added to SSVEP epochs for the following classification analysis.
\subsection{Accuracy and ITR on Various Algorithms}
In this study, eleven frequency recognition methods, including six supervised methods and five training-free methods, are adopted to evaluate the BETA database. For S1-S15, 2-s length of epochs is used for analysis and for S16-S70 the epoch length is 3 s. A sliding window from the stimulus onset (latency corrected) with an interval of 0.2 s is applied to epochs for offline analysis. 
\subsubsection{Supervised methods}
\begin{figure}[h!]
\begin{center}
\includegraphics[width=18cm]{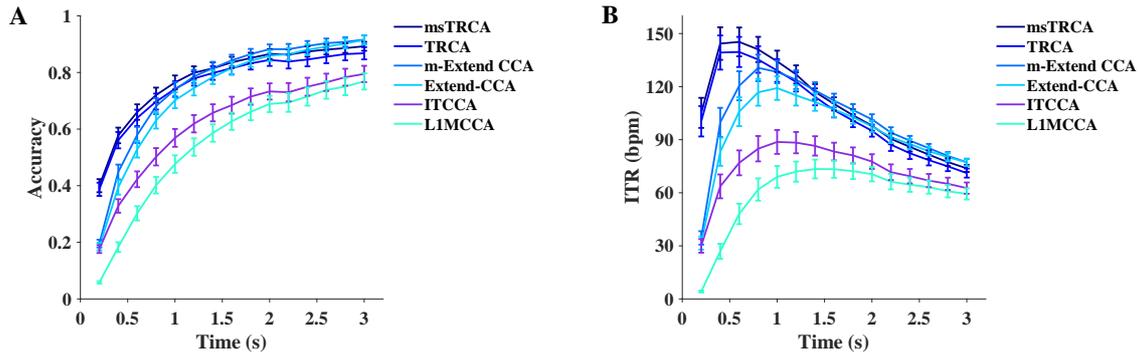}
\end{center}
\caption{Average classification accuracy \textbf{(A)} and ITR \textbf{(B)} for 6 supervised methods (msTRCA, TRCA, m-Extended CCA, Extended CCA, ITCCA and L1MCCA). Ten data lengths ranging from 0.2 s to 3 s with an interval of 0.2 s are used for evaluation. The gaze shift time is 0.55 s for the calculation of ITR.}\label{fig:8}
\end{figure}

 We choose six supervised methods including task-related component analysis (TRCA, \cite{nakanishi2018enhancing}), multi-stimulus task-related component analysis (msTRCA, \cite{wong2019learning}), Extended CCA (\cite{nakanishi2014a}), modified Extended CCA (m-Extended CCA, \cite{chen2015high-speed}), L1-regularized multiway CCA (L1MCCA, \cite{zhang2013l1-regularized}) and individual template-based CCA (IT-CCA, \cite{bin2011a}) for comparison. To calculate accuracy and ITR, a leave-one-out procedure on four blocks is applied to each subject. Figure \ref{fig:8} illustrates the average accuracy (A) and ITR (B) for supervised methods. The result shows that msTRCA outperforms other methods in data lengths less than 1.4 s and m-Extended CCA achieves the highest performance in data lengths from 1.6 s to 3 s. One-way repeated measures ANOVA reveals that there are significant differences between these methods in accuracy and ITR for all time windows. Specifically, for short time window at 0.6 s, post hoc paired t-tests show that msTRCA \textgreater TRCA \textgreater m-Extended CCA \textgreater Extended CCA \textgreater ITCCA \textgreater L1MCCA in accuracy and ITR. Here '\textgreater' indicates $p$ is less than 0.05 in ITR with Bonferroni correction for pairwise comparison between the two sides. For a medium-length time window at 1.4 s, post hoc paired t-tests show that m-Extended CCA \textgreater msTRCA \textgreater Extended CCA / TRCA \textgreater ITCCA \textgreater L1MCCA (Extended CCA vs TRCA: $p=1$ in ITR; Bonferroni corrected). The data length of highest ITR varies between different methods (msTRCA: 145.26 $\pm$ 8.15 bpm at 0.6 s, TRCA: 139.58 $\pm$ 8.52 bpm at 0.6 s, m-Extended CCA: 130.58 $\pm$ 7.53 bpm at 0.8 s, Extended CCA: 119.17 $\pm$ 6.67 bpm at 1 s, ITCCA: 88.72 $\pm$ 6.75 bpm at 1 s, L1MCCA: 73.42 $\pm$ 5.31 bpm at 1.4 s).
 
\subsubsection{Training-free methods}
\begin{figure}[h!]
\begin{center}
\includegraphics[width=16cm]{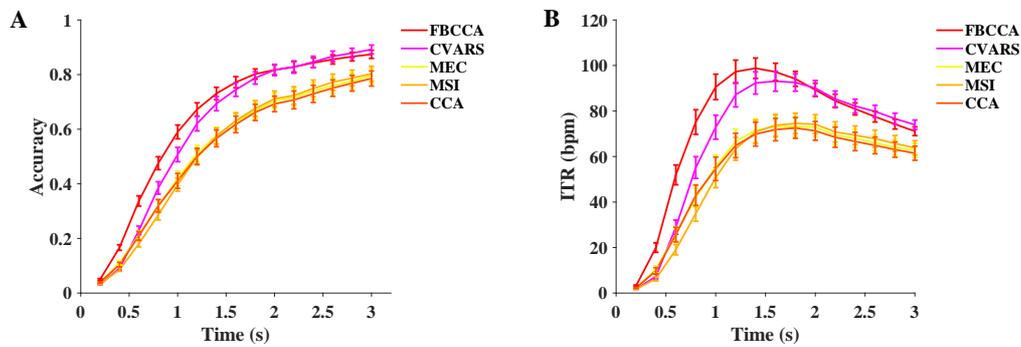}
\end{center}
\caption{Average classification accuracy \textbf{(A)} and ITR \textbf{(B)} for 5 training-free methods (FBCCA, CVARS, MEC, MSI and CCA). Ten data lengths ranging from 0.2 s to 3 s with an interval of 0.2 s are used for evaluation. The gaze shift time is 0.55 s for the calculation of ITR.}\label{fig:9}
\end{figure}

In this study, we compare five training-free methods, including minimum energy combination (MEC, \cite{friman2007multiple}), canonical correlation analysis (CCA, \cite{bin2009an}), multivariate synchronization index (MSI, \cite{zhang2014multivariate}), filter bank canonical correlation analysis (FBCCA, \cite{chen2015filter}) and canonical variates with autoregressive spectral analysis (CVARS, \cite{Abu2016Advancing}). As illustrated in Figure \ref{fig:9}, FBCCA is superior over other methods in data lengths less than 2 s and CVARS outperforms others from 2s to 3s. Significant differences are found between these methods in accuracy and ITR by one-way repeated measures ANOVA for all data lengths. Post hoc paired t-tests show that FBCCA \textgreater CVARS \textgreater CCA / MSI / MEC in ITR for medium data length at 1.4s, $p \textless 0.05$ (Bonferroni corrected) for all pairwise comparisons except CCA vs MSI ($p=1$), CCA vs MEC ($p=1$), MSI vs MEC ($p=1$) in ITR. For training-free methods, the highest ITR is achieved after 1.2 s (FBCCA: 98.79 $\pm$ 4.49 bpm at 1.4 s, CVARS: 93.08 $\pm$ 4.39 bpm at 1.6 s, CCA: 72.54 $\pm$ 4.54 bpm at 1.8 s, MSI: 74.54 $\pm$ 4.46 bpm at 1.8 s,  MEC: 73.23 $\pm$ 4.43 bpm at 1.8 s).

 \par Note that for TRCA and msTRCA, the ensemble and filter-bank scheme are employed as the default. For fair comparison, the number of harmonics $N_h$ is set 5 in all the methods with sinusoidal templates except the m-Extended CCA according to \cite{chen2015high-speed} ($N_h=2$). For all the methods without filter bank scheme, trials are band-pass filtered between 6 Hz and 80 Hz in line with the previous study (\cite{nakanishi2015comparison}), except for the CVARS method.
\subsection{Correlation between ITR and SNR}
\begin{figure}[h!]
\begin{center}
\includegraphics[width=14cm]{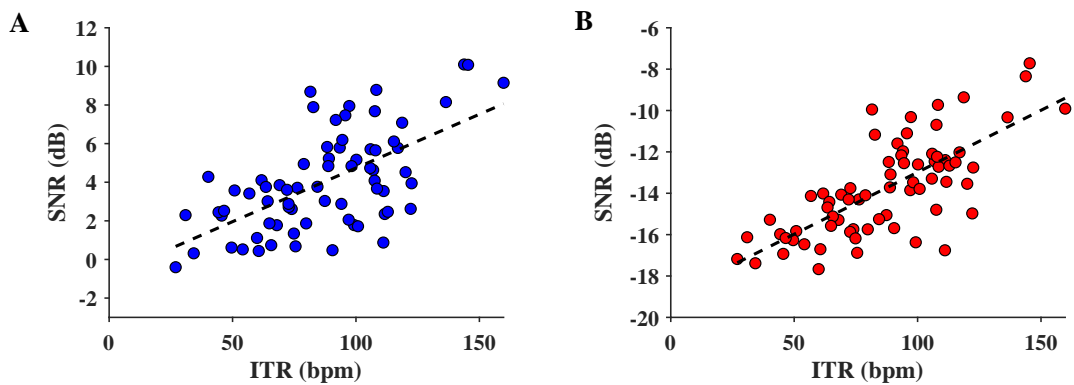}
\end{center}
\caption{The scatter plot of narrow-band SNR vs ITR \textbf{(A)} and wide-band SNR vs ITR \textbf{(B)}. The dash line indicates a linear model regressed on the data (\textbf{A}: $R^2$=0.368, p\textless0.001; \textbf{B}: $R^2$=0.531, p\textless0.001). The regression indicates wide-band SNR correlate better with ITR than narrow-band SNR.}\label{fig:10}
\end{figure}

To investigate the relationship between the SNR and ITR metrics, we plot the scatter diagram of ITR with SNR for wide band and narrow band, respectively. The maximum ITR for each subject (after averaging by block) from the training-free FBCCA is chosen for analysis. Figure \ref{fig:10} illustrates the scatter plot of narrow-band SNR vs ITR (A) and wide-band SNR vs ITR (B). As can be seen from the figure, ITR is positively correlated with SNR, either narrow-band or wide-band. We further fit the data with a linear model and the adjusted R squared ($R^2$) that measures the goodness-of-fit can be obtained. The result reveals that an adjusted $R^2$ of 0.368 for the narrow-band SNR (p\textless0.001) and an adjusted $R^2$ of 0.531 for the wide-band SNR (p\textless0.001). This indicates that the metric of wide-band SNR is more correlated with and can better predict ITR than narrow-band SNR.
\subsection{BCI quotient}
\begin{figure}[h!]
\begin{center}
\includegraphics[width=10cm]{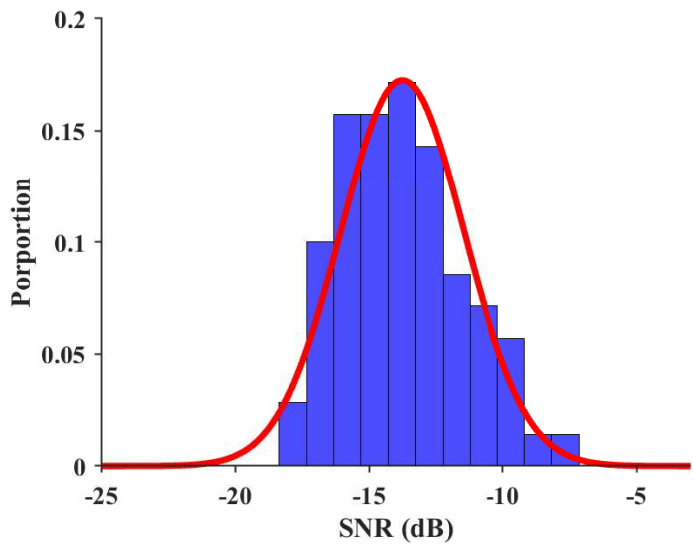}
\end{center}
\caption{The distribution of wide-band SNR and its fit by a normal distribution. An individual's SNR of SSVEP is rescaled to the range of normal distribution in equation (7) to obtain the BCI quotient.}\label{fig:11}
\end{figure}

Electroencephalographic signals including SSVEP show individual differences in population. In this study, we propose a BCI quotient to characterize the subject's capacity to use SSVEP-BCI measured at a population level. Equivalent to the scoring procedure of intelligence quotient (IQ) (\cite{wechsler2008wechsler}), the (SSVEP-) BCI quotient is defined as follows
\begin{equation}
\rm{Quotient_{BCI}}=15\cdot\frac{SNR-\mu}{\sigma}+100
\end{equation}
where SNR represents the wide-band SNR, mean $\mu=-13.76$ and standard deviation $\sigma=2.31$ in this study (Figure \ref{fig:11}). The mean and standard deviation can be estimated more accurately for a larger database in the future. The BCI quotient rescales an individual's SNR of SSVEP to a range of normal distribution $\mathcal{N}(100,15)$. Since it is a relative value from SNR and SNR is correlated with ITR, the BCI quotient has the potential to measure signal quality and performance for individuals in SSVEP-BCI. Higher values of BCI quotient indicate a higher probability of good BCI performance. For instance, the BCI quotients of S20 and S23 are 74.62 and 139.25, respectively, which reveals a prior the individual level of ITR (59.91 bpm for S20 and 145.41 bpm for S23).
\section{Discussion}
\subsection{Data quality and its applicability}
Compared to the benchmark database (\cite{wang2017a}), the BETA database has lower SNR and corresponding ITR in classification (for benchmark database: FBCCA, 117.96 $\pm$ 7.78 bpm at 1.2 s; m-Extended CCA, 190.41 $\pm$ 7.90 bpm at 0.8 s; CCA, 90.16 $\pm$ 6.81 bpm at 1.6s; 0.55-s rest time for comparison (\cite{wang2017a,chen2015filter})). This is reasonable since in BCI applications there is actually no electromagnetic shielding condition and neither can we ensure that each subject has a high SNR of SSVEP. The discrepancy in SNR is also due in part to the distinct stimulus duration, which is 2s or 3s for the BETA database and 5s for the benchmark database. But even taking the same stimulus duration in to consideration (3-s trial after stimulus onset for 55 subjects in the BETA and 35 subjects in the benchmark), the BETA database is also significantly lower in SNR (narrow-band SNR: BETA 4.319$\pm$0.021 dB, benchmark 5.239$\pm$0.020 dB, $p \textless 0.05$; wide-band SNR: BETA -13.510$\pm$0.015 dB, benchmark -12.650$\pm$0.015 dB, $p \textless 0.05$). Therefore, the present BETA database poses challenges for traditional frequency recognition methods and also opens up opportunities for the development of robust frequency recognition algorithms in real-world applications. A large number of subjects in BETA has the merit of reducing over-fitting and can provide an unbiased estimation in the evaluation of algorithms. Also, the large volume of BETA provides the substrate for the study of transfer learning to exploit the common discriminative patterns across subjects. Note that the number of blocks of each subject is smaller than that in the benchmark database. Since reducing the training and calibration time is critical for the BCI application, the proposed database can serve as the test-bed for the development of supervised frequency recognition methods based on smaller training samples or few-shot learning. It is noteworthy that the application scenario of BETA database is not limited to the 40-target speller in the study. Practitioners can select a subset of the 40 targets (e.g. 4, 8, 12 targets) and design customized paradigms to suit the need in a variety of real-world applications. With the advent of big data, the BETA has the potential to facilitate modeling the brain at a population level and help develop novel classification approaches or learning methodology, e.g. federated learning (\cite{mcmahan2016communication}) based on the big data. 
\subsection{Supervised and training-free methods}
In general, the state-of-the-art supervised frequency recognition methods have the advantage of higher performance in ITR and the training-free methods excel in ease of use. In this study, two of the supervised methods (m-Extended CCA, Extended CCA) outperform the five training-free algorithms for all data lengths. Specifically, for the short-time window (0.2-1 s) the supervised methods (msTRCA, TRCA, m-Extended CCA, Extended CCA) outperform the training-free methods by a large margin (Supplementary Figure 1). This is because the introduction of EEG training template and learned spatial filters facilitates SSVEP classification. For time window longer than 2 s (2.2-3 s), post hoc paired t-tests show that no significant difference is between m-Extended CCA and Extended CCA, between FBCCA and CVARS, and among ITCCA, CCA, MEC and MSI (p\textgreater 0.05, Bonferroni corrected). This suggests some common mathematical grounds shared by these algorithms in principle (\cite{wong2020spatial}). Interestingly, the TRCA method drops in performance as reported in the previous study (\cite{nakanishi2018enhancing}), presumably caused by the lack of sufficient training block for subjects with low SNR. As evidenced by the previous study (\cite{nakanishi2018enhancing}), for TRCA the number of training data greatly affects classification accuracy ($\approx$ 0.85 with 11 training blocks and $\approx$ 0.65 with two training blocks for 0.3-s time window). This implies that methods with a sinusoidal reference template (e.g. m-Extended CCA, Extended CCA and FBCCA etc.) may be more robust than methods without it. To sum up, the classification analysis in the present study demonstrates the utility of different competing methods on BETA. The comparison of different methods on a single database complements the previous work of \cite{zerafa2018train}, where the performance of various methods is not compared on the same database. 
\subsection{SNR comparison}
In the SNR analysis, we find the wide-band SNR more correlated with ITR compared to the narrow-band SNR. From Figure \ref{fig:5}, it is worth noting that a transition from narrow-band SNR to wide-band SNR does not change the relative relationship between the SNRs of two databases. Nevertheless, the wide-band SNR metric reduces skewness of data distribution (from -0.719 to -0.096 for benchmark database; from -1.089 to -0.142 for BETA; narrow-band SNR followed by wide-band SNR), which renders the SNR statistic favorably more Gaussian in distribution. Since the spectral power of a signal is equal to its power in the time domain according to the Parseval's Theorem, the formulated wide-band SNR has equivalent mathematical underpinning as the metric of temporal SNR counterpart. Apart from its expressive power of wide-band SNR, the metric is also intuitive in the description of signal and noise due to the frequency tagging attribute of SSVEP. 

\section{Concluding remark}
\par Here we present a novel \textbf{BE}nchmark database \textbf{T}owards BCI \textbf{A}pplication (BETA) for the 40-target SSVEP-BCI paradigm. The BETA database is featured by its large number of subjects and its paradigm that is well-suited for real-world applications. The quality of the BETA is validated by the typical temporal, spectral and spatial profile of SSVEP, together with the SNR and the estimated visual latency. On the BETA database, we compared eleven frequency recognition methods, including 6 supervised methods and 5 training-free methods. The result of the classification analysis validates the data and demonstrates the performance of different methods in one arena as well. As for the metric to characterize SSVEP, we recommend adopting the wide-band SNR at the single-trial level and use the BCI quotient at the population level. We expect the BETA database would be a test-bed for the development of method and paradigm for practical BCI and push the boundary of BCI toward real-world application.

\section*{Conflict of Interest Statement}
The authors declare that the research was conducted in the absence of any commercial or financial relationships that could be construed as a potential conflict of interest.

\section*{Author Contributions}

B.L. conducted the data curation, analysis and wrote the manuscript. X.H. designed the paradigm and performed the data collection. Y.W and X.C. performed the data collection and revised the manuscript. X.G. supervised the study.

\section*{Funding}
B.L is supported by the Doctoral Brain$+$X Seed Grant Program of Tsinghua University. This research is supported by the National Natural Science Foundation of China under grant 61431007, and the National Key R\&D Program of China under grant 2017YFB1002505, and Guangdong Province, China under grant 2018B030339001.

\section*{Acknowledgments}
We would like to thank D.X., J.S., L.L., X.L. for providing support in the data collection.




\section*{Figure captions}


\begin{figure}[h!]
\begin{center}
\includegraphics[width=20cm]{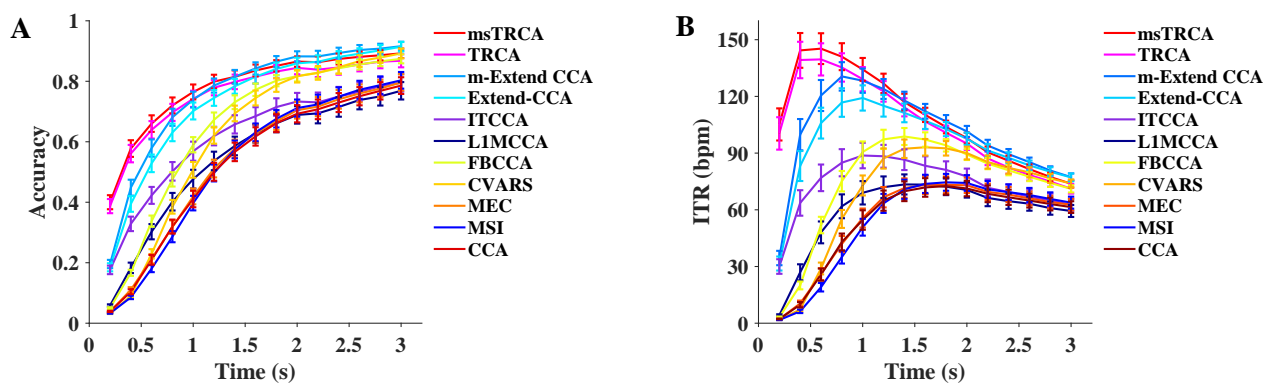}
\end{center}
\caption{(Suppl.) Average classification accuracy \textbf{(A)} and ITR \textbf{(B)} for 11 frequency recognition methods (msTRCA, TRCA, m-Extended CCA, Extended CCA, ITCCA, L1MCCA, FBCCA, CVARS, MEC, MSI and CCA). Ten data lengths ranging from 0.2 s to 3 s with an interval of 0.2 s are used for evaluation. The gaze shift time is 0.55 s for the calculation of ITR.}
\end{figure}

\end{document}